\documentclass[10pt, conference, compsocconf]{IEEEtran}
\usepackage{algorithmicx}
\usepackage[ruled]{algorithm2e}
\usepackage{times,epsfig}
\usepackage{epstopdf}
\usepackage{epsfig}
\usepackage{subfigure}
\usepackage{multirow}
\usepackage{graphicx}
\usepackage{setspace}
\usepackage{algpseudocode}
\usepackage{amsmath}
\usepackage{graphics}
\usepackage{amsmath,bm}
\usepackage{color}
\usepackage{cite}
\usepackage{mathrsfs}

\ifCLASSINFOpdf
\else
\fi

\begin{document}
%
\title{Layered Image Compression using Scalable Auto-encoder}

\author{Chuanmin Jia$^{1,2}$, Zhaoyi Liu$^{1,3}$, Yao Wang$^1$, Siwei Ma$^2$, Wen Gao$^2$\\
\thanks{This work was done when C. Jia and Z. Liu were visiting students at Tandon School of Engineering, New York University.}
$^1$Department of Electrical and Computer Engineering, Tandon School of Engineering, \\
New York University, Brooklyn, NY 11201, USA\\
$^2$Peking University, Institute of Digital Media, Beijing 100871, China\\
$^3$Beijing Institute of Technology, School of Information and Electronics, Beijing 100081, China\\
yw523@nyu.edu, \{cmjia, swma, wgao\}@pku.edu.cn, 20091305@bit.edu.cn
}

\maketitle

\begin{abstract}
  This paper presents a novel convolutional neural network (CNN) based image compression framework via scalable auto-encoder (SAE).
   Specifically, our SAE based deep image codec consists of hierarchical coding layers, each of which is an end-to-end optimized auto-encoder.
   The coarse image content and texture are encoded through the first (base) layer while the consecutive (enhance) layers iteratively code the pixel-level reconstruction errors between the original and former reconstructed images.
   The proposed SAE structure alleviates the need to train  multiple models for different bit-rate points by recently proposed auto-encoder based codecs.
   The SAE layers can be combined to realize multiple rate points, or to produce a scalable stream.
   The proposed method has similar rate-distortion performance in the low-to-medium rate range as the state-of-the-art CNN based image codec (which uses different optimized networks to realize different bit rates) over a standard public image dataset. Furthermore, the proposed codec generates better perceptual quality in this bit rate range.
\end{abstract}


\IEEEpeerreviewmaketitle

\section{Introduction}
Image compression aims at representing an image with minimal coding bits while preserving the maximal pixel-level reconstruction quality as it could be.
Recently, deep learning (DL) based image compression has been one of the emerging topics due to its elegant end-to-end optimization ability.
Multiple learning-based image codecs have been proposed by investigating the joint intersection of deep learning and image coding.
Essentially, the deep models are trained to learn the image-to-image mapping between the pristine image and the reconstructed image based on the rate-distortion (R-D) learning objective.

Different from conventional image coding formats, JPEG~\cite{wallace1992jpeg}, JPEG2000~\cite{skodras2001jpeg} and BPG~\cite{bellard2015bpg} (based on High Efficiency Video Coding, HEVC~\cite{sullivan2012overview}), which utilize separate sub-modules for prediction and transform coding, the deep codecs formulate the end-to-end learnt networks as transform coding~\cite{goyal2001theoretical}. The most representative models typically adopted the auto-encoder (AE) like structure, which generate latent representations that are quantized and entropy coded. For example, Toderici~\textit{et al.}~\cite{toderici2017full} proposed an end-to-end image coding model based on the recurrent neural network (RNN), where the original and residual images are iteratively compressed using RNN structure.
During each RNN iteration, the better reconstruction with a commensurate bitrate (2-bits) cost will be produced. However, the RNN based method might have limitations in representing high-frequency residuals. Additionally, the lack of explicit entropy estimation during RNN training also constraints its overall R-D performance.
To address this issue, Theis~\textit{et al.}~\cite{theis2017lossy} presented a convolutional neural network (CNN) based AE structure, where the entropy model was approximated using Gaussian distribution during optimization. In~\cite{baig2017learning}, an inpainting based learning approach was proposed for image compression. To enhance the visual quality, the generative adversarial network (GAN) based learning strategies were embeded in the CNN based framework~\cite{agustsson2018generative,rippel2017real} to improve the perceptual quality of the reconstructed images. In~\cite{balle2016end}, the generalized divisive normalization (GDN)~\cite{balle2015density} was introduced as a substitute for the nonlinear activation in a variational autoencoder (VAE) to de-correlate the channel-wise dependency among latent representations, which significantly improves the coding performance to be competitive with JPEG2000 standard. Compared with other activation functions, the core advantage of GDN is its full reversibility, which guarantees nearly no information loss for the transform coding. More recently~\cite{balle2018variational}, a novel distribution parameter estimation method was proposed for entropy coding which has brought additional coding gain.

All the aforementioned approaches have to train a particular auto-encoder for each target bit rate, which can limit their applicability when the desired bit rates have to be adapted in realtime, and/or when it is impractical to save multiple trained models. Inspired by the conventional scalable coding paradigm~\cite{ohm2005advances}, we proposed a scalable auto-encoder (SAE) based deep image coding method to solve this problem by iteratively and incrementally coding the errors using the end-to-end trained auto-encoders. By cascading the bitstreams generated by each layer of SAE, variable bit-rates or layered bit streams could be obtained while maintaining optimal R-D performances. Furthermore, the proposed SAE structure could be compatible with any other learning based image codec since each layer of SAE could be substituted by different AE based deep codecs.

\section{Proposed Method}\label{proposed}
\begin{figure}[!htb]
\center
\includegraphics[width=0.41\textwidth, height=0.38\textheight]{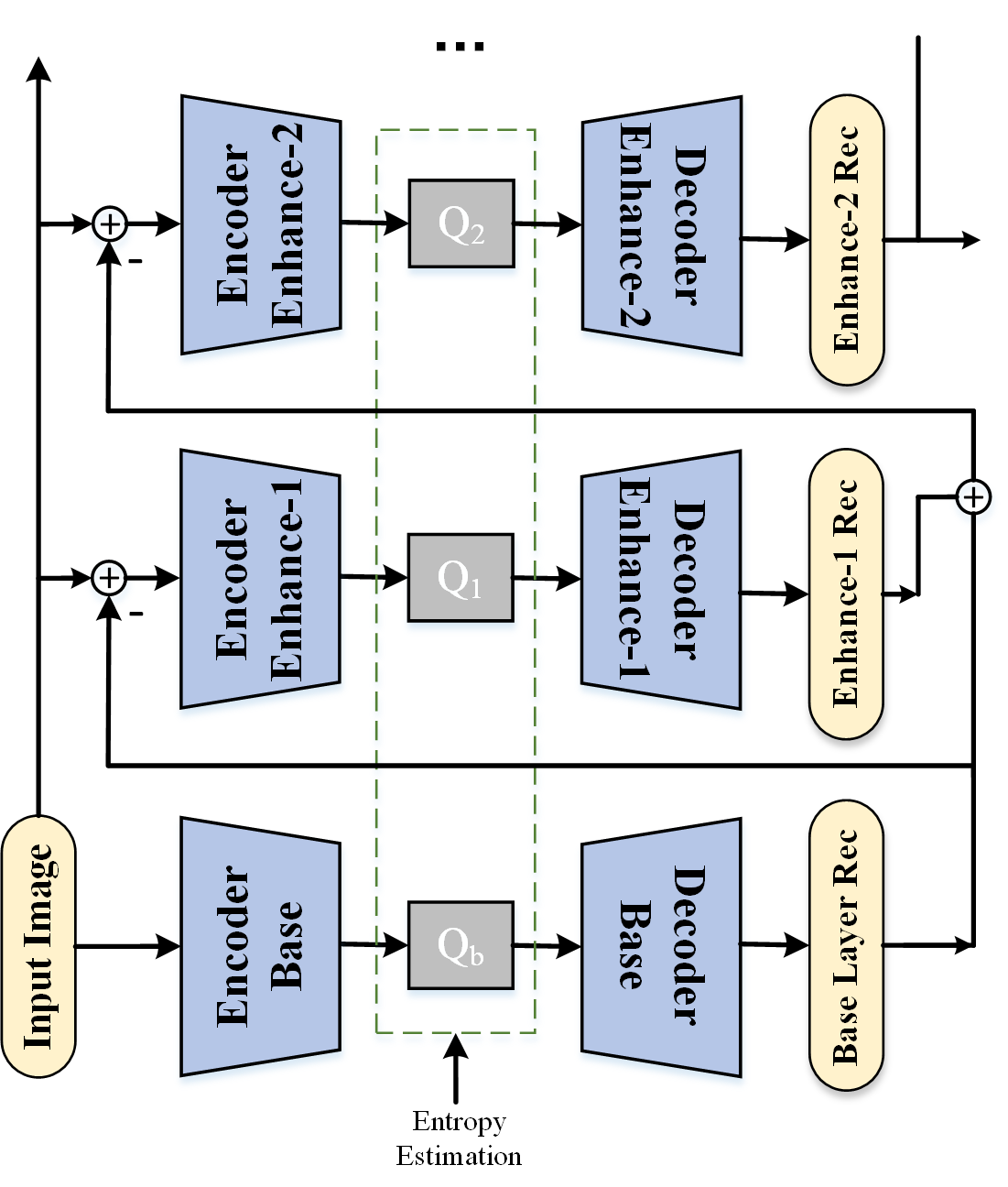}
\footnotesize
\caption{ The framework of the proposed SAE based image compression (first three layers are illustrated).
}
\label{Fig:framework}
\end{figure}

The overall flowchart of the proposed framework is depicted in Fig.~\ref{Fig:framework}. This model contains several stacked modules, each of which is an end-to-end optimized AE.
The original image is firstly compressed by the AE in the base layer. Subsequently, the enhance layers would take the difference between the latest reconstructed image and the original image as input then compress the residue. We train our entire framework in a layer-by-layer manner, which means we fix all previous layers when training the current one. The design of base layer and enhance layers will be introduced in more details in subsequent subsections.

\subsection{Base Layer}
\begin{figure}[!htb]
\center
\includegraphics[width=0.4\textwidth]{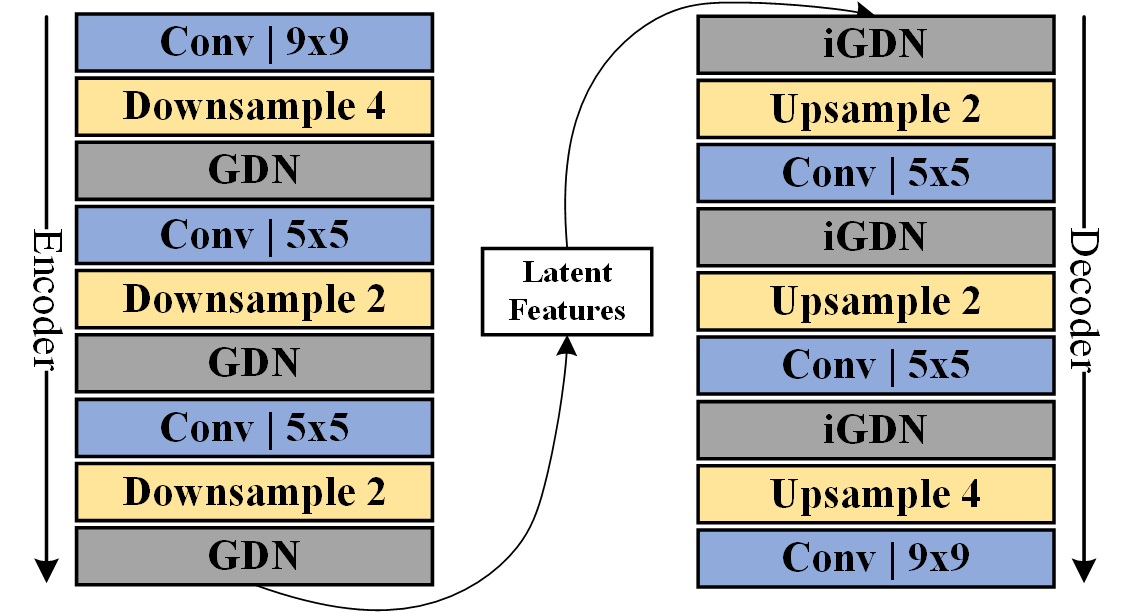}
\footnotesize
\caption{The AE in each layer of our proposed SAE. The encoder and decoder are shown in the left and right panels respectively.
}
\label{Fig:autoencoder}
\end{figure}

The base layer in the proposed SAE mainly compresses the coarse image content and basic texture information.
Considering the original image $x$, the encoder part of AE in the base layer encodes $x$ into latent representation $q_{b}$.
\begin{equation} \label{Eq:base-enc}
q_{b}=E_{b}(x),
\end{equation}
Subsequently, the quantizer ($round(\cdot)$) is applied to the latent representation $q_{b}$ to obtain the quantized latent feature $\bar{q_{b}}$.
The bits required to code $\bar{q_{b}}$ can be approximated by its entropy, which is determined from $P_{\bar{q_{b}}}$, which is the marginal probability mass function of $\bar{q_{b}}$. Following~\cite{balle2016end}, for the purpose of end-to-end differentiability of the loss function, we replace the quantizer by adding a white noise with uniform distribution in (-1/2, 1/2) to the latent feature $q_{b}$.
We considered two different training objectives. The first one is aimed to optimize the objective quality measured by the mean square error, and is trained using the following loss function,
\begin{equation} \label{Eq:base-obj}
Loss_{b}=||x-{\hat x}_{b}||^{2}_{2} + \lambda_{b} * R(q_{b}+\Delta q),
\end{equation}
where $\lambda_{b}$ is the Lagrange multiplier and ${\hat x}_{b}=D_{b}(\bar{q_{b}})$ is the reconstructed image of base layer and $D_{b}(\cdot)$ represents the base layer decoder of AE (shown in Fig.~\ref{Fig:autoencoder}).
The second training objective optimizes for perceptual quality measured by MS-SSIM~\cite{wang2003multi}, which is
\begin{equation} \label{Eq:base-obj-msssim}
Loss_{b}=(1-{\rm MS\_SSIM}({\hat x}_{b})) + \lambda_{b_{msssim}} * R(q_{b}+\Delta q),
\end{equation}
where ${\rm MS\_SSIM}(\cdot)$ is based on~\cite{wang2003multi} and $\lambda_{b_{msssim}}$ is the Lagrange multiplier for MS-SSIM.
To estimate the rate $R(\cdot)$, we deploy the state-of-the-art entropy model described in~\cite{balle2018variational}.

Following~\cite{balle2016end}, the $E_{b}(\cdot)$ has three convolution layers with filter size $9\times9$, $5\times5$ and $5\times5$ with the down-sampling step size $4, 2, 2$ respectively. GDN is utilized to achieve non-linearity after each convolution layer. The decoder ($D_{b}(\cdot)$) has the mirror structure of $E(\cdot)$. The AE structure is depicted in Fig.~\ref{Fig:autoencoder},  which is the same as~\cite{balle2016end}. Note that the same structure is used in each subsequent enhance layer and the only difference is the number of channels for the latent features, which will be described later.

\subsection{Enhance Layers}
In the proposed framework, the enhance layers are responsible for iteratively encoding the residues between the reconstructed image from the previous layers and the original image.
As shown in Fig.~\ref{Fig:framework}, the first enhance layer (enhance-1) takes the error between original image $x$ and the reconstructed image of base layer as input.
The formulation of enhance-1 could be represented as follows.
\begin{equation} \label{Eq:enhance1}
\tilde{x}_{e1}=D_{e1}(round(E_{e1}(x-\hat{x}_{b}))),
\end{equation}
To acquire the reconstruction from base layer and enhance-1, we could simply add the two outputs from such two layer ($\hat{x}_{e1}=\hat{x}_{b}+\tilde{x}_{e1}$). As such, the reconstruction quality could be enhanced with the error coded in the enhance layer.

For the subsequent enhance layers, taking the second enhance layer (enhance-2) as example, the input is the residue between original $x$ and the reconstruction from the latest layer $\hat{x}_{e1}$. The reconstructed residue can be described as,
\begin{equation} \label{Eq:enhance2}
\tilde{x}_{e2}=D_{e2}(round(E_{e2}(x-\hat{x}_{e1}))),
\end{equation}
In this case, the reconstruction for enhance-2 is obtained by adding three corresponding outputs of the three AEs ($\hat{x}_{e2}=\hat{x}_{b}+\tilde{x}_{e1}+\tilde{x}_{e2}$).

The loss function of training the $i$-th enhance layer has the following formation,
\begin{equation} \label{Eq:enhance-obj}
Loss_{ei}=||x-\hat{x}_{ei}||^{2}_{2} + \lambda_{ei} * R(q_{ei}+\Delta q),
\end{equation}
where the $\lambda_{ei}$ is the Lagrange Multiplier for $i$-the layer.
Similar with base layer training, uniform noise is also added to the latent variables $q_{ei}$ to train the enhance layers for relaxation of the loss functions.
To obtain rate-distortion optimality, we tried different combination of $\lambda$'s, which are detailed in the next Section.

\section{Training Details}\label{trainingdetails}
This section presents the training details of the proposed SAE based image codec.
A subset of ImageNet~\cite{krizhevsky2012imagenet} database is used for training, which contains 5500 RGB images. And another 300 images are used for validation.
The convergence is met when the loss on the validation images becomes stable.
We randomly crop a region of $256\times256$ from each training sample to prevent boundary issues.
The hyper-parameters during our training procedure are listed in Table.~\ref{tab:hyper-param}.
For each SAE layer, we need to train a pair of encoder and decoder as well as a entropy model, iteratively.
The $MSE$~$lr$ means the fixed learning rate for the AE and entropy model in Eq.~(\ref{Eq:base-obj}) and $Rate$~$lr$ is the initialized learning rate of the entropy model.
The $Rate$~$lr$ has an exponential decay with the parameter 0.96 for every $5,000$ iterations during training.
$Adam$~\cite{kingma2014adam} is utilized as the optimizer for both AE and entropy model respectively.

We train each layer successively: given the previous ($i$-1) layers, we try to find the optimal hyper-parameters for the $i$-th SAE layer,
including the number of feature maps and $\lambda$, that achieve the best rate-distortion tradeoff.
\begin{table}[t]
  \centering
  \renewcommand{\arraystretch}{1.2}\footnotesize
  \caption{The hyper-parameters for training SAE.}
  \begin{tabular}{c|c}
  \hline
  \hline
  {\bf Parameter Name}     &{\bf Value} \\ \hline
  MSE lr         &0.0001   \\ \hline
  Rate lr        &0.001    \\ \hline
  Optimizer      &$Adam$     \\ \hline
  epochs         &1000     \\ \hline
  batch size     &8        \\ \hline
  training image size     &$256\times256$        \\ \hline \hline
  \end{tabular}
  \label{tab:hyper-param}
\end{table}

\subsection{Base Layer Training}
To achieve a good rate-distortion tradeoff for the base layer, and to use as few channels as possible for reduced complexity,
we varied the number of feature maps in each of the three convolution layers in the encoder and decoder of the AE and $\lambda$ values.
We found that using 48 features maps in all three layers, and $\lambda$=3000 achieved the best result for the MSE-oriented optimization.
For MS-SSIM-oriented optimization, $\lambda$=50 achieved the best result.

\subsection{Enhance Layer Training}
We train each subsequent enhance layers, while fixing the lower layers at their optimized states.
We have found that the number of features for enhance layers should increase with the growing of enhance layers. We suspect that this is because the distribution of the errors become more similar to random noise when the enhance layer goes deeper such that more parameters are needed to model and capture such distribution. The parameter $\lambda$, which is responsible for balancing the contributions from the entropy term and the MSE term in Eq.~(\ref{Eq:enhance-obj}), should decrease layer by layer since more emphasis should be put on minimizing the MSE. Recall that ideally $\lambda$ should be equal to the negative slope of the MSE vs. rate curve, and this slope reduces as the rate increases. Table.~\ref{tab:lambda-fmaps} summarizes the $\lambda$ values and the number of features for each layer in the trained SAE.
These values are selected based on exhaustive search of different combinations of the parameters when training the SAE structure.
\begin{table}[t]
  \centering
  \renewcommand{\arraystretch}{1.2}\footnotesize
  \caption{$\lambda$ and number of feature maps for each layer of the proposed SAE.}
  \begin{tabular}{c|c|c|c|c|c}
  \hline \hline
  Layers                      &Base & e1    &e2        & e3  & e4 \\ \hline
  $\lambda$ (for MSE)         &3000 &1000   & 300	   &100  & 30    \\ \hline
  $\lambda$ (for MS-SSIM)     &50   &30     & 10	   &0.5  & -   \\ \hline
  Feature Map Number          &48   &48     & 96 	   &144  & 192  \\ \hline \hline
  \end{tabular}
  \label{tab:lambda-fmaps}
\end{table}

\subsection{Entropy Model}
In this paper, we re-use the entropy model proposed in~\cite{balle2018variational}, which incorporates a hyper-prior to effectively capture spatial dependencies in the latent representation generated by each layer of the proposed SAE. Particularly, we deploy different entropy models for different layers in the proposed SAE.

\section{Experimental Results}\label{experiments}
For training and testing, we used the popular DL library Tensorflow~\cite{abadi2016tensorflow} and the Tensorflow-compression submodule~\cite{tfc2018}, which is an implementation of~\cite{balle2018variational}.

\subsection{Experiment Set-up}
To evaluate the efficiency of the propose SAE based image codec, we test the proposed model on the widely used Kodak Lossless True Color Image dataset~\cite{kodak1999} who contains 24 true color images with resolution $512\times768$. The results presented in this section are the average of these 24 images. It is worthy noting that all the experiments and comparisons are based on three-channel true color images.
The test environment of this work is the Intel i5 7200U-CPU with 16GB RAM and NVIDIA GTX 1050Ti GPU.

\subsection{Rate-distortion Performances}
We compare our proposed SAE coder against the algorithms in~\cite{balle2016end} and~\cite{balle2018variational}.
Our method and~\cite{balle2016end} have the same structure for the encoder and decoder, as illustrated in Fig.~\ref{Fig:autoencoder}. And~\cite{balle2018variational} used one more convolution layer on top of~\cite{balle2016end}. However, the numbers of feature maps differ among these methods.~\cite{balle2016end} used 192 feature maps in each layer while~\cite{balle2018variational} used 128 and 192 for the convolution layer and bottleneck layer at low bit-rate and that of 192, 320 for high bit-rate points respectively. The numbers of feature maps in the proposed SAE differ among the scalable layers and are summarized in Table.~\ref{tab:lambda-fmaps}.

\begin{figure}[!htb]
\center
\includegraphics[width=0.47\textwidth, height=0.3\textheight]{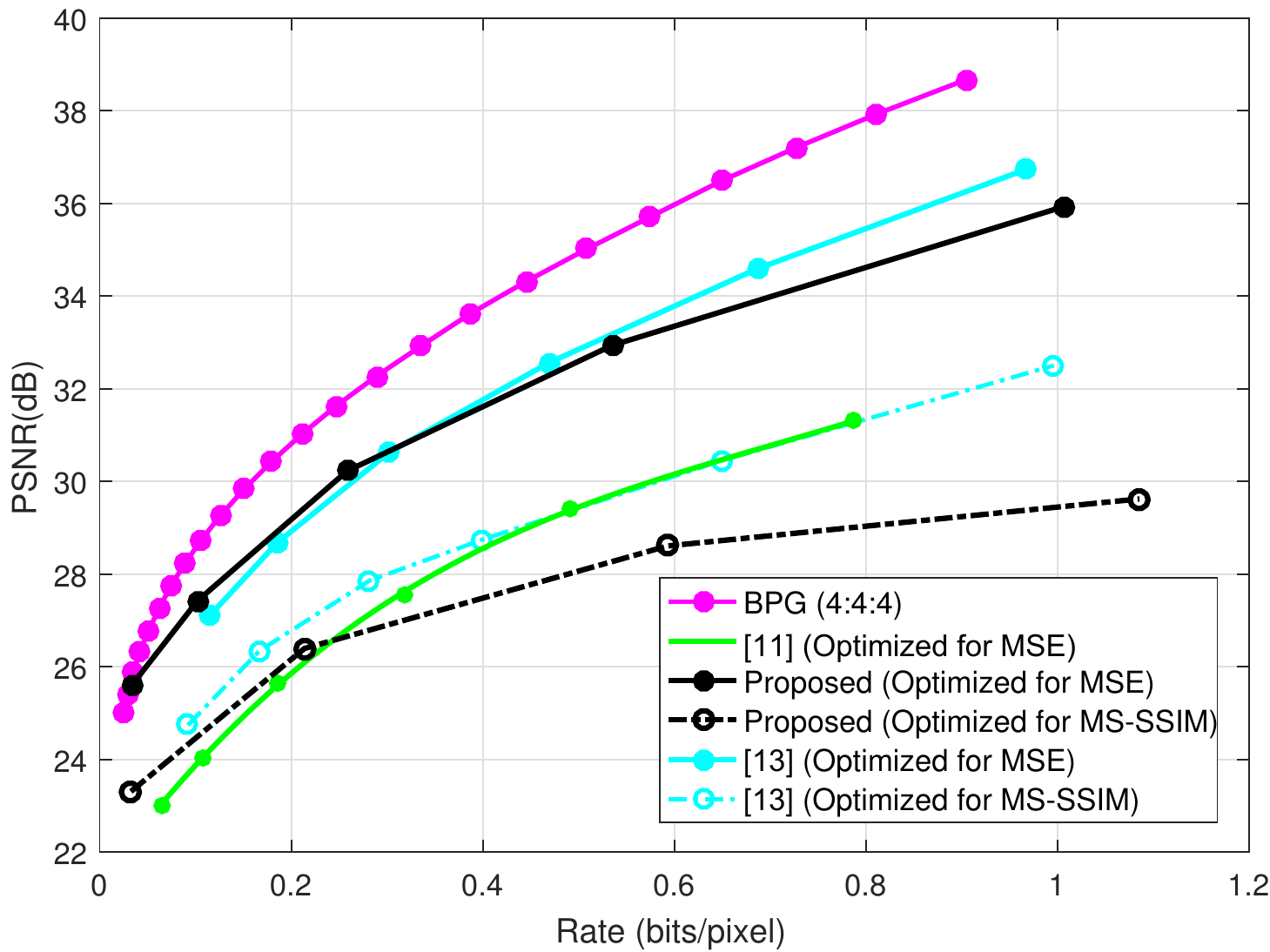}
\footnotesize
\caption{R-D curves (PSNR) over Kodak dataset.
}
\label{Fig:PSNR}
\end{figure}

\begin{figure}[!htb]
\center
\includegraphics[width=0.47\textwidth, height=0.3\textheight]{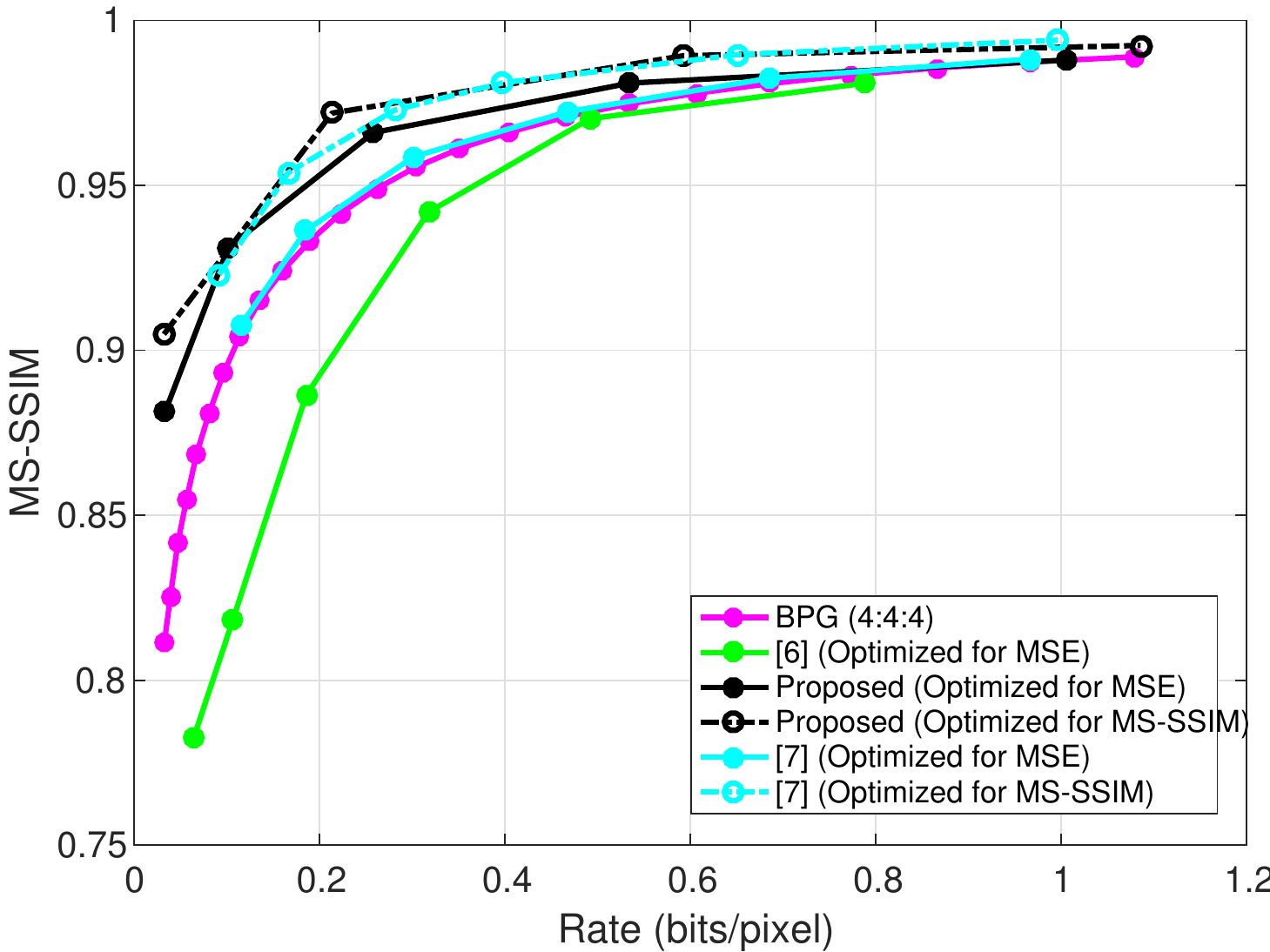}
\footnotesize
\caption{R-D curves (MS-SSIM) over Kodak dataset.
}
\label{Fig:MSSSIM}
\end{figure}

The proposed method use the same entropy estimation method as~\cite{balle2018variational}, but~\cite{balle2016end} used an older method, which is less efficient.
\begin{table}[t]
  \centering
  \renewcommand{\arraystretch}{1.2}\footnotesize
  \caption{The R-D performance of the proposed SAE based image codec.}
  \begin{tabular}{c|c|c|c|c}
  \hline \hline
  \multicolumn{1}{c|}{\multirow{2}{*}{Dataset}}	&\multicolumn{2}{c|}{vs.~\cite{balle2016end}} &\multicolumn{2}{c}{vs.~\cite{balle2018variational}}\\ \cline{2-5}%
	        & BD-rate        & BD-PSNR         & BD-rate  & BD-PSNR    \\ \hline
  Kodak 	  & -65.2 \%       & 3.38 dB          & -0.6 \%  & 0.021 dB          \\ \hline \hline
  \end{tabular}
  \label{tab:rdperformance}
\end{table}
To illustrate the coding performance in a wide range of bit-rates,
the R-D curves of the proposed SAE,~\cite{balle2016end} and~\cite{balle2018variational} are provided in Fig.~\ref{Fig:PSNR}.
We also provide the R-D curves obtained by the BPG codec$^{1}$. For each of~\cite{balle2018variational} and SAE, we provide two sets of results, one optimized for MSE and another for MS-SSIM.
Table III summarizes the BD-Rate and BD-PSNR of the proposed SAE method against~\cite{balle2016end} and~\cite{balle2018variational}, respectively.
Compared with~\cite{balle2016end}, the SAE achieved significant coding gain, where over 65\% bit-rate could be reduced and 3.38dB PSNR increase could be obtained.
The SAE performance is slightly better than~\cite{balle2018variational}.
\footnotetext[1]{Note that when using the BPG codec~\cite{bellard2015bpg}, we set the option to indicate that  the input image is in the RGB format. The BPG performance reported in~\cite{balle2018variational} is lower than in Fig.~\ref{Fig:PSNR} and~\ref{Fig:MSSSIM}, because the option was set to assume the input is in YCbCr format.}

For the proposed SAE model, the points from the left to the right correspond to the results of the base layer, enhance-1 layer to enhance-4 layer. Obviously, both~\cite{balle2018variational} and the proposed model outperforms~\cite{balle2016end} over the entire range of bit-rate by clear margin.
This is due to the more efficient entropy coding method used.  The SAE coder is similar to~\cite{balle2018variational} up to about 0.5 bpp, and then becomes less efficient. This loss of coding efficiency with more layers is as expected, as with any scalable coder compared to a non-scalable coder.
In fact, it was somewhat surprising that SAE was able to achieve similar performance (in fact slightly better) as~\cite{balle2018variational}, up to 3 enhancement  layers.

Additionally, we provide the comparisons based on MS-SSIM in Fig.~\ref{Fig:MSSSIM}.
In general, the MS-SSIM metric is better correlated with the perceptual quality than PSNR.
It is very encouraging that the proposed SAE method has similar or better performance than~\cite{balle2018variational} in the entire rate range.
Moreover, the SAE method achieved much better performance than BPG in the entire bit-rate range, consistent with the visual evaluation described below.

\subsection{Visual Evaluations}

\begin{figure}[]
\center
\begin{minipage}[]{0.99\textwidth}

\subfigure[~\cite{balle2016end}: 0.1455 bpp, 24.1610 dB, 0.8827]{
\includegraphics[width=0.16\textwidth, height=0.15\textheight]{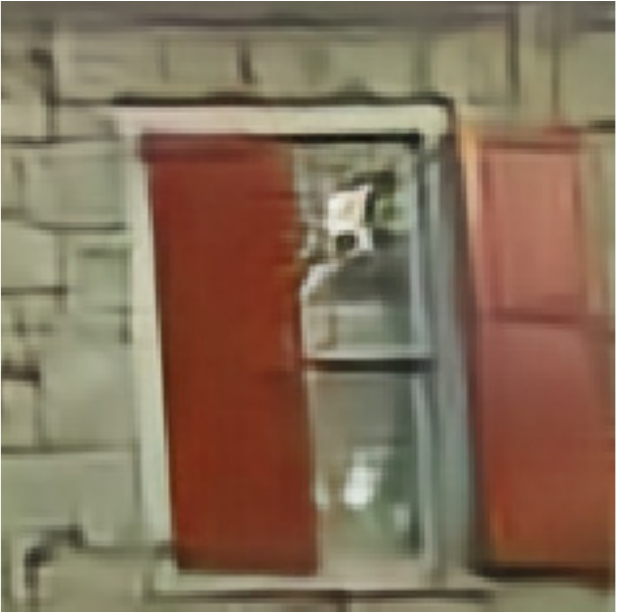}}
\subfigure[~\cite{balle2018variational}: 0.1782 bpp, 25.0016 dB, 0.9079]{
\includegraphics[width=0.16\textwidth, height=0.15\textheight]{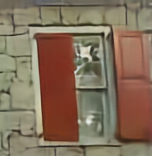}}
\subfigure[Ours: 0.1305 bpp, 24.9626 dB, 0.9250]{
\includegraphics[width=0.16\textwidth, height=0.15\textheight]{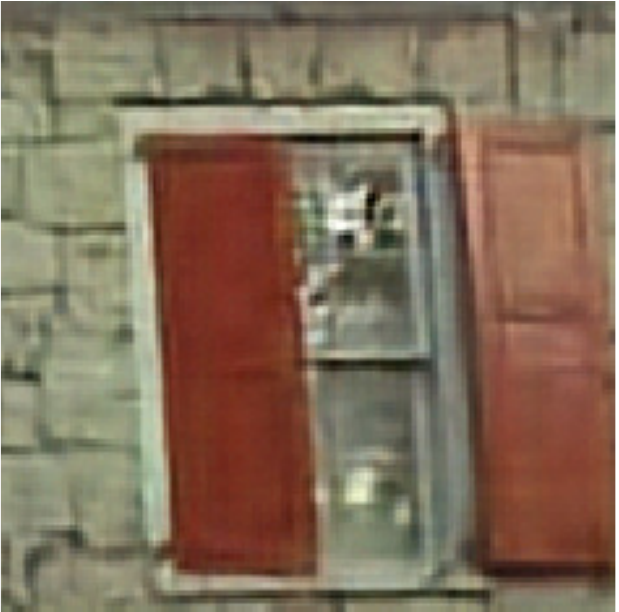}}
\vfill
\subfigure[~\cite{balle2016end}: 0.1394 bpp, 25.1619 dB, 0.8739]{
\includegraphics[width=0.16\textwidth, height=0.15\textheight]{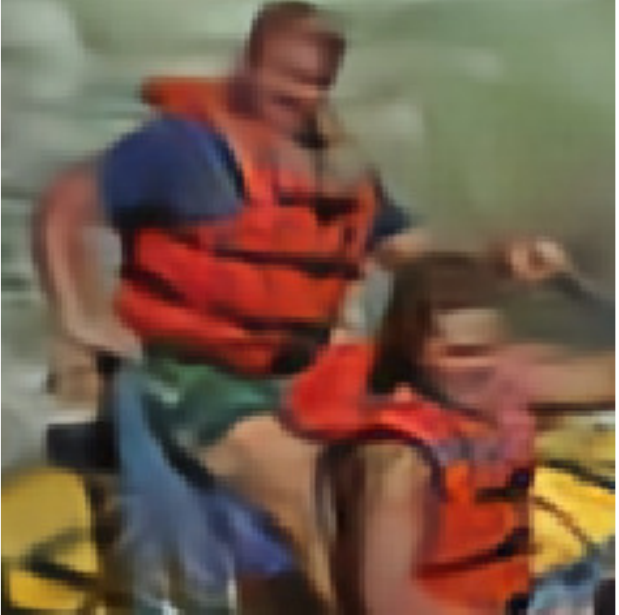}}
\subfigure[~\cite{balle2018variational}: 0.1417 bpp, 25.5955 dB, 0.8870]{
\includegraphics[width=0.16\textwidth, height=0.15\textheight]{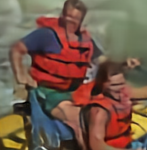}}
\subfigure[~Ours: 0.1166 bpp, 25.8482 dB, 0.9192]{
\includegraphics[width=0.16\textwidth, height=0.15\textheight]{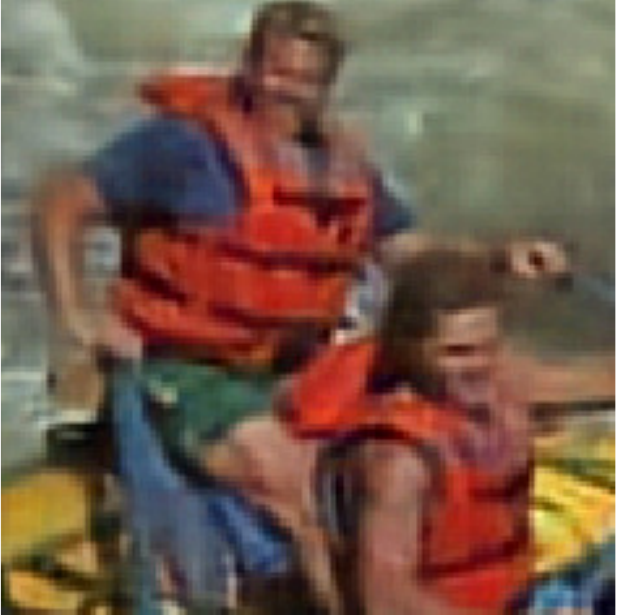}}
\label{Fig.main}
\end{minipage}

\begin{minipage}[]{0.99\textwidth}
\subfigure[~\cite{balle2016end}: 0.6468 bpp, 28.9845 dB, 0.9725]{
\includegraphics[width=0.16\textwidth, height=0.15\textheight]{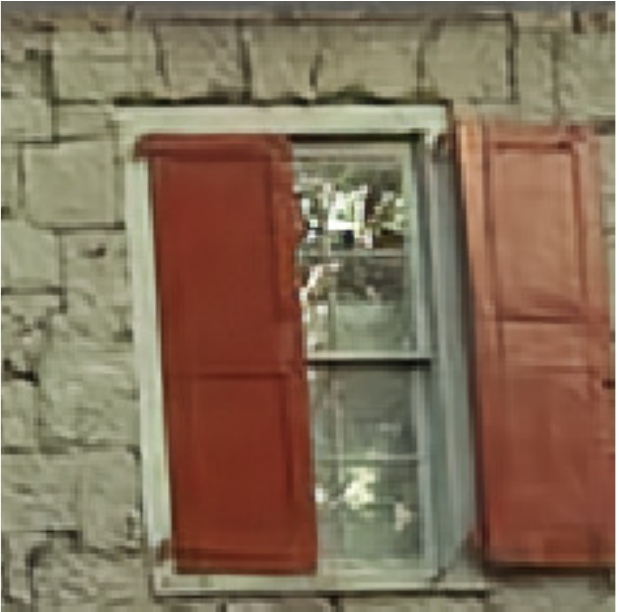}}
\subfigure[~\cite{balle2018variational}: 0.4959 bpp, 28.3934 dB, 0.9648]{
\includegraphics[width=0.16\textwidth, height=0.15\textheight]{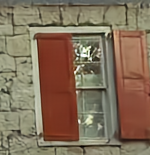}}
\subfigure[Ours: 0.5495 bpp, 30.9506 dB, 0.9852]{
\includegraphics[width=0.16\textwidth, height=0.15\textheight]{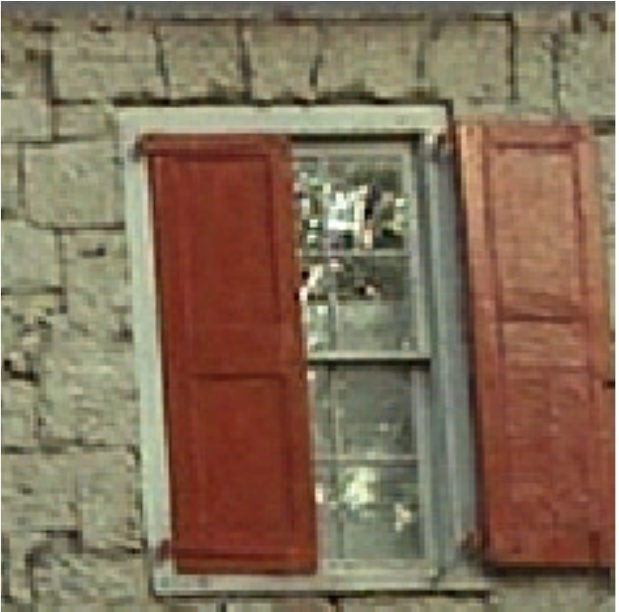}}
\vfill
\subfigure[~\cite{balle2016end}: 0.5782 bpp, 30.0684 dB, 0.9680]{
\includegraphics[width=0.16\textwidth, height=0.15\textheight]{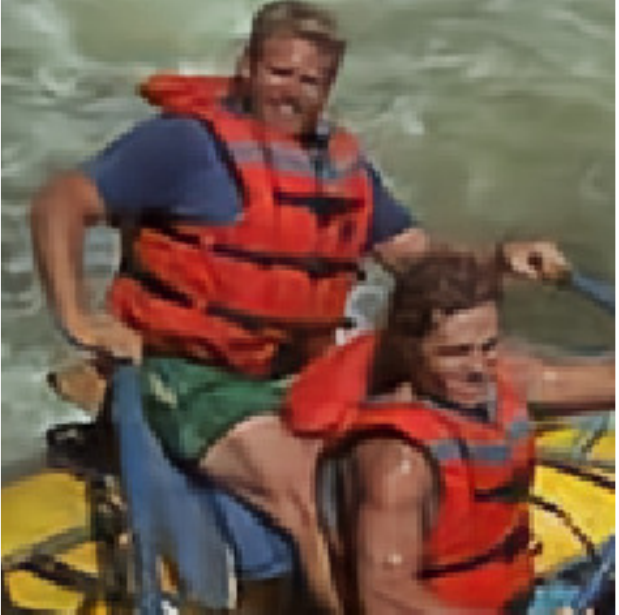}}
\subfigure[~\cite{balle2018variational}: 0. 4858 bpp, 29.1530 dB, 0.9533]{
\includegraphics[width=0.16\textwidth, height=0.15\textheight]{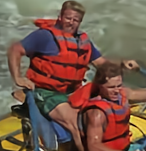}}
\subfigure[~Ours: 0.4932 bpp, 28.7907 dB, 0.9831]{
\includegraphics[width=0.16\textwidth, height=0.15\textheight]{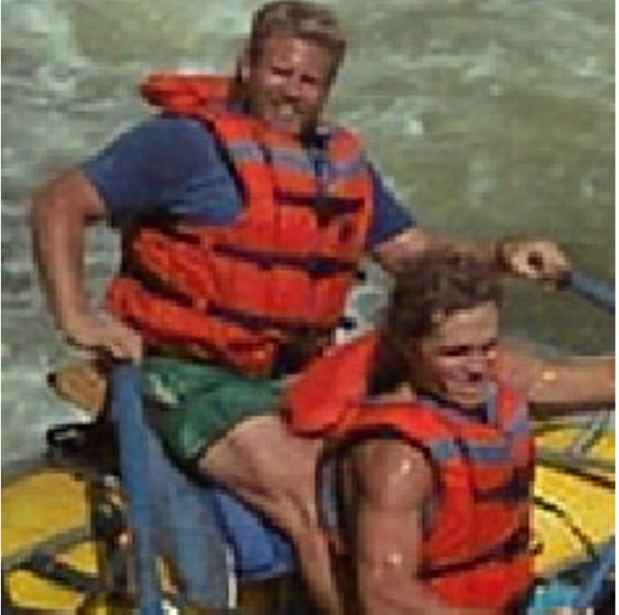}}
\label{Fig.main}
\end{minipage}
\caption{ Comparisons of decoded test images by~\cite{balle2016end},~\cite{balle2018variational} and proposed method. The last number under each figure is the MS-SSIM value.}
\label{Fig:visual-compare}
\end{figure}

\begin{figure}[t]
\center

\begin{minipage}[]{0.99\textwidth}
\subfigure[~\cite{balle2016end}: 0.1128 bpp, 26.2665 dB, 0.8831]{
\includegraphics[width=0.14\textwidth, height=0.1\textheight]{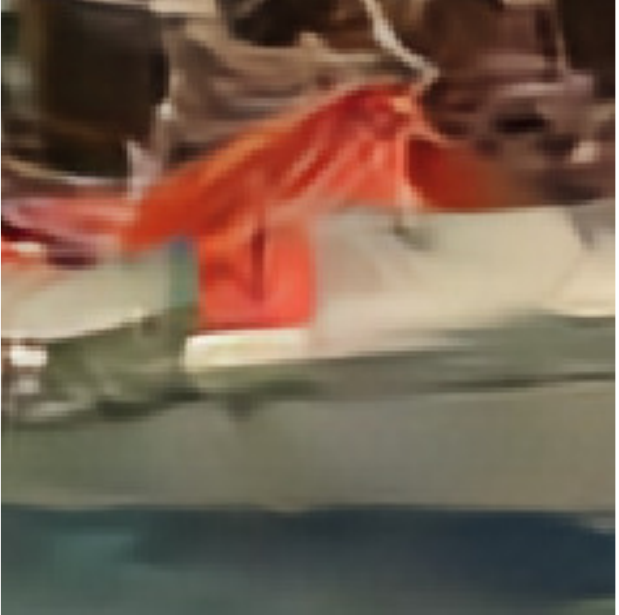}}
\subfigure[~\cite{balle2018variational}: 0.1102 bpp, 30.0579 dB, 0.9483]{
\includegraphics[width=0.14\textwidth, height=0.1\textheight]{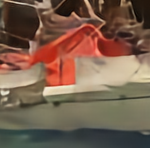}}
\subfigure[Ours: 0.0996 bpp, 27.0657 dB, 0.9228]{
\includegraphics[width=0.14\textwidth, height=0.1\textheight]{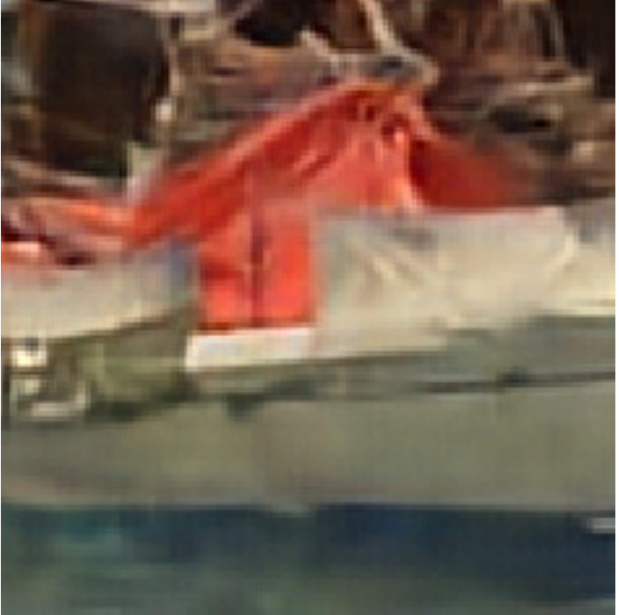}}
\vfill
\subfigure[~\cite{balle2016end}: 0.0850 bpp, 29.9462 dB, 0.9361]{
\includegraphics[width=0.14\textwidth, height=0.1\textheight]{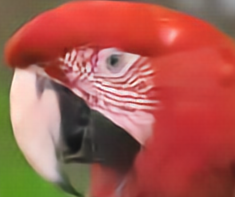}}
\subfigure[~\cite{balle2018variational}: 0.0739 bpp, 30.1352 dB, 0.9388]{
\includegraphics[width=0.14\textwidth, height=0.1\textheight]{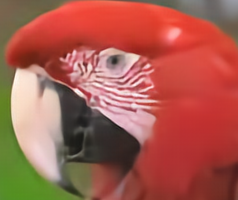}}
\subfigure[Ours: 0.0811 bpp, 30.6522 dB, 0.9524]{
\includegraphics[width=0.14\textwidth, height=0.1\textheight]{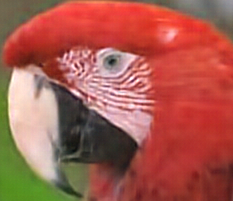}}

\subfigure[~\cite{balle2016end}: 0.4858 bpp, 30.9572 dB, 0.9644]{
\includegraphics[width=0.14\textwidth, height=0.1\textheight]{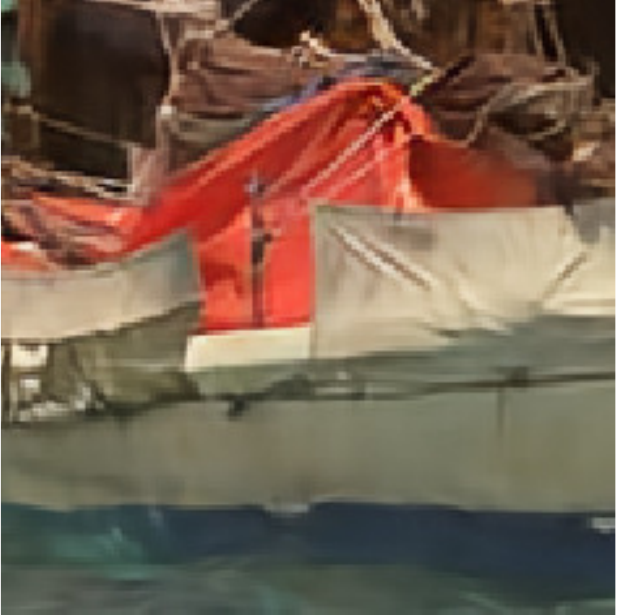}}
\subfigure[~\cite{balle2018variational}: 0.4596 bpp, 31.9486 dB, 0.9660]{
\includegraphics[width=0.14\textwidth, height=0.1\textheight]{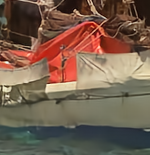}}
\subfigure[Ours: 0.4336 bpp, 32.6423 dB, 0.9809]{
\includegraphics[width=0.14\textwidth, height=0.1\textheight]{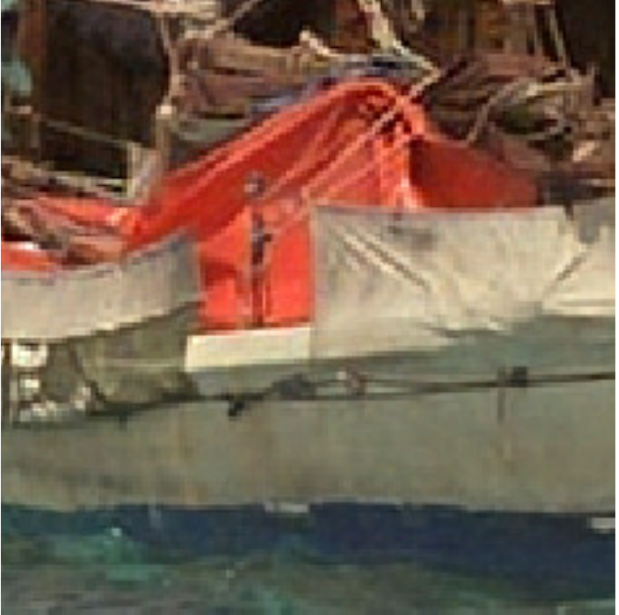}}
\vfill
\subfigure[~\cite{balle2016end}: 0.1370 bpp, 24.0123 dB, 0.8881]{
\includegraphics[width=0.14\textwidth, height=0.1\textheight]{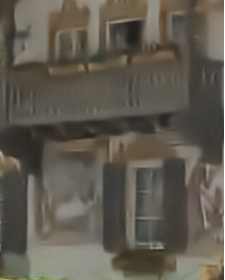}}
\subfigure[~\cite{balle2018variational}: 0.1387 bpp, 24.4523 dB, 0.9004]{
\includegraphics[width=0.14\textwidth, height=0.1\textheight]{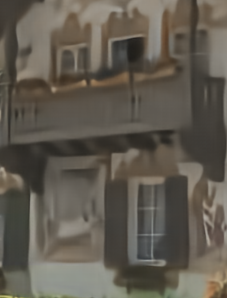}}
\subfigure[Ours: 0.1130 bpp, 24.6560 dB, 0.9287]{
\includegraphics[width=0.14\textwidth, height=0.1\textheight]{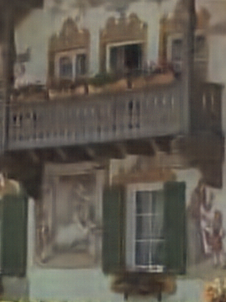}}

\end{minipage}
\caption{ Comparisons of decoded test images by~\cite{balle2016end},~\cite{balle2018variational} and proposed method. The last number under each figure is the MS-SSIM value.}
\label{Fig:visual-compare2}
\end{figure}

For visual comparisons, Fig.~\ref{Fig:visual-compare} and Fig.~\ref{Fig:visual-compare2} present several cropped versions of reconstructed images from Kodak dataset. Notice that the proposed SAE structure offers more detailed information in contour and textural regions while using less or similar bits than both~\cite{balle2016end} and~\cite{balle2018variational} at the low to intermediate bit rate.
We have provided all of the decoded test images by the proposed SAE framework  and methods of~\cite{balle2016end,balle2018variational} in different bit-rate points in the supplementary materials$^{1}$.
\footnotetext[1]{Please visit {\texttt{https://github.com/chuanminj/MIPR2019/}}. \\ for supplementary materials.}

\section{Conclusion}
In this paper, a scalable auto-encoder based deep image codec is proposed.
The novelty of the paper lies in that the proposed method does not need to train multiple independent networks to realize different bit-rate points.
The quantitative and qualitative evaluations have shown that the proposed method can achieve rate-distortion performance similar to a state-of-art DL-based method in the low to intermediate bit rate in terms of mean-square error, and has similar or better performance than the benchmark in terms of the perceptual quality in the entire rate range.
We should also note that the proposed SAE structure is general. One can simply replace the particular AE structure in Fig.~\ref{Fig:autoencoder} by another structure that can provide better coding performance in each layer. In fact, one can also use methods not based on AEs.

\section*{Acknowledgment}
The authors would like to thank Dr. Johannes Ball{\'e} for kindly providing their trained models of~\cite{balle2016end} and the decoded images of~\cite{balle2018variational} for performance comparison. This work was done by C. Jia and Z. Liu as visiting students in NYU-Tandon sponsored by China Scholarship Council (CSC), which is gratefully acknowledged.

%
%

\bibliographystyle{IEEEtran}
\bibliography{IEEEabrv,mipr}

\end{document}